\documentclass[aps,prd,twocolumn,superscriptaddress,floatfix,nofootinbib,natbib,longbibliography]{revtex4-1}
\usepackage{graphicx,longtable,mathrsfs,color,array}
\usepackage[usenames,dvipsnames]{xcolor} 
\usepackage{amssymb,amsmath,mathtools,mathrsfs,slashed} 
\usepackage{epsfig,subfigure,placeins,float} 
\usepackage{booktabs,longtable,ctable,multirow} 
\usepackage{exscale,relsize} 
\usepackage[normalem]{ulem} 
\usepackage{enumerate}
\usepackage{enumitem}
\usepackage[colorlinks,linkcolor=blue,citecolor=blue,urlcolor=blue ]{hyperref}

\makeatletter
\@fpsep\textheight
\makeatother

\usepackage[stable]{footmisc}
\usepackage{color}

\begin{document}
\title{Reconstruction of the neutrino mass as a function of redshift}
\author{Christiane S. Lorenz}
\email{chrlorenz@phys.ethz.ch}
\affiliation{Institute for Particle Physics and Astrophysics, ETH Zürich, Wolfgang-Pauli-Strasse 27, CH-8093 Zürich, Switzerland}
\author{Lena Funcke}
\email{lfuncke@perimeterinstitute.ca}
\affiliation{Perimeter Institute for Theoretical Physics, 31 Caroline Street North, Waterloo, Ontario, N2L 2Y5, Canada}
\author{Matthias Löffler}
\affiliation{Seminar for Statistics, Department of Mathematics, ETH Zürich, Rämistrasse 101, CH-8092 Zürich, Switzerland}
\author{Erminia Calabrese}
\affiliation{School of Physics and Astronomy, Cardiff University, The Parade, Cardiff, CF24 3AA, United Kingdom}

\date{Received \today; published -- 00, 0000}

\begin{abstract}
We reconstruct the neutrino mass as a function of redshift, $z$, from current cosmological data using both standard binned priors and linear spline priors with variable knots. Using 
cosmic microwave background temperature, polarization and lensing data, in combination with distance measurements from baryonic acoustic oscillations and supernovae, we find that the neutrino mass is consistent with $\sum m_\nu(z)=$ const.
We obtain a larger bound on the neutrino mass at low redshifts coinciding with the onset of dark energy domination, $\sum m_\nu(z=0)<1.46$~eV (95\% CL). This result can be explained either by the well-known degeneracy between $\sum m_\nu$ and $\Omega_\Lambda$ at low redshifts, or by models in which neutrino masses are generated very late in the Universe. We finally convert our results into cosmological limits for models with non-relativistic neutrino decay and find $\sum m_\nu <0.21$~eV (95\% CL), which would be out of reach for the KATRIN experiment.
\end{abstract}

\maketitle

\section{Introduction}
Cosmological surveys and particle physics experiments are independent and complementary probes of neutrino properties. Neutrino oscillation experiments have measured the squared mass differences between neutrino mass eigenstates, giving a lower bound of 59~meV for the total sum of the neutrino masses, $\sum m_\nu$~\cite{10.1093/ptep/ptaa104}. In addition, the Karlsruhe Tritium Neutrino Experiment (KATRIN) has constrained the electron neutrino mass to be lower than $m_{\nu,e}<0.8$~eV at 90\% CL~\cite{Aker:2021gma,Aker:2021ibz}. Independent from these constraints, measurements of the cosmic microwave background (CMB) with the \textit{Planck} satellite mission, combined with distance measurements from baryonic acoustic oscillations (BAO) from the Sloan Digital Sky Survey (SDSS) and 6dF have provided a tight upper bound on the total sum of neutrino masses, $\sum m_\nu<120$~meV at 95\% CL~\cite{Aghanim:2018eyx}.

Contrary to neutrino mass direct detection limits, the cosmological neutrino mass bound assumes a specific cosmological model, usually the $\Lambda$Cold Dark Matter ($\Lambda$CDM) model or its single-parameter extensions. In more extended cosmological models, the cosmological neutrino mass limits can become less stringent due to opening wider parameter spaces and/or covering more complex physics scenarios.

On the one hand, the \textit{lower} bound of $\sum m_\nu=59$~meV imposed by neutrino oscillation experiments can be relaxed in cosmological analyses to $\sum m_\nu=0$~meV, for example, if cosmological neutrinos disappear in the late Universe. Although the original ``neutrinoless Universe'' proposal~\cite{Beacom2004} has been ruled out\footnote{By free-streaming of the cosmic neutrino background before photon decoupling \cite{Hannestad2004,Lancaster2017}, by the resulting phase shift in the CMB peaks~\cite{Follin:2015hya}, and by precise CMB measurements of the effective number of species in the early Universe~\cite{Aghanim:2018eyx}.}, other scenarios like models predicting postrecombination neutrino mass generation and subsequent relic neutrino annihilation~\cite{Dvali:2016uhn} are still possible. In such models, the neutrino mass parameter cannot be captured using cosmological data and can only be measured using terrestrial and astrophysical experiments such as KATRIN. 
The possible cosmological disappearance of the neutrino mass parameter has also been proposed in the context of modified gravity theories~\cite{Bellomo:2016xhl,Hagstotz:2019gsv,Hagstotz2019}; these should soon be tested with surveys like Euclid~\cite{Hagstotz2019}. If we allow for strong fine tuning, another possibility to eliminate the cosmological neutrino mass bounds would be to postulate a new light scalar particle that couples to neutrinos, with a coupling constant that needs to be smaller than $g \sim 10^{-7}$ to avoid laboratory constraints~\cite{esteban2021long}. 

On the other hand, the \textit{upper} cosmological neutrino mass bound is sensitive to a number of model assumptions and can be slightly relaxed, for example when the dark energy equation of state is allowed to vary in time~\cite{2012PhRvD..86b3526J,PhysRevD.87.083523,Allison:2015qca,Lorenz:2017fgo,Yang:2017amu,Vagnozzi:2018jhn,Choudhury:2018byy,RoyChoudhury:2019hls}, when the curvature of the Universe is not fixed~\cite{RoyChoudhury:2019hls}, when considering additional relativistic degrees of freedom~\cite{Aghanim:2018eyx}, or when assuming non-standard momentum distributions of the cosmic neutrinos~\cite{Oldengott:2019lke}. Moreover, it has been shown that the cosmological neutrino mass bound can be substantially weakened when neutrinos are unstable and thus their lifetime is smaller than the age of the Universe~\cite{Escudero2019,Chacko:2019nej,Chacko:2020hmh,Escudero:2020ped,Barenboim:2020vrr} or when neutrino masses are varying in time
~\cite{Fardon:2003eh,Dvali:2016uhn,Koksbang:2017rux,Lorenz:2018fzb}. Note, however, that the strongly relaxed neutrino mass bounds of $\sum m_\nu$$<0.9$~eV (95\% CL) with neutrino decays~\cite{Chacko:2019nej} or $\sum m_\nu$$<4.8$~eV (95\% CL) with time-varying neutrino masses~\cite{Lorenz:2018fzb} have been derived from previous releases of cosmological data,
including the \textit{Planck} 2015 CMB 
data. The neutrino mass bound has been substantially tightened with the \textit{Planck} 2018 release, 
$\sum m_\nu$=$ 0.12$ eV (95\% CL, \textit{Planck} 2018 TTTEEE+lowE+lensing+BAO~\cite{Aghanim:2018eyx}) and therefore, we expect the above-mentioned bounds also to change. 

In a previous publication~\cite{Lorenz:2018fzb}, some of us investigated how the standard cosmological neutrino mass bound would be affected if neutrino masses were to be generated late in the Universe. This analysis suggested that the combination of current CMB temperature, polarization and lensing data, as well as BAO and supernovae (SN) data prefers neutrino masses to be generated at low redshifts, allowing a significantly larger cosmological neutrino mass upper bound. An alternative interpretation of the results is that current cosmological datasets do not necessarily require neutrinos to be massive, pushing the scale factor of the phase transition in the model to very low redshifts. The trend of larger neutrino mass bounds in the late Universe had already been noted before (see, e.g., Refs.~\cite{Brookfield_2006,Brookfield2006,Koksbang:2017rux,Battye:2013xqa,Wyman:2013lza,Beutler:2014yhv,Poulin:2018zxs} and references therein). This could either arise due to the well-known degeneracy between neutrino masses and dark energy, or be explained by new physics beyond the Standard Model of Cosmology. In particular, it seems striking that the energy scales of dark energy and neutrino masses are numerically very close, $\sqrt[4]{\rho_\Lambda}\sim m_\nu\sim{\rm meV}$. If cosmological data permit larger neutrino masses during dark energy domination, there could be an intriguing theoretical connection between these two phenomena (see, e.g., Refs.~\cite{Fardon:2003eh,Bjaelde:2007ki,Ayaita:2014una,Mandal:2019kkv,DAmico:2018hgc,Dvali:2016uhn} and references therein).

The tight neutrino mass limit from cosmology is making a direct detection from Tritium $\beta$-decay experiments like KATRIN very challenging. Indeed, KATRIN's target sensitivity of 200 meV at 90\% CL for the $\nu_e$ mass~\cite{Drexlin:2013lha} will unlikely hit the cosmological limit obtained in either $\Lambda$CDM or its simple  extensions~\cite{Aghanim:2018eyx,RoyChoudhury:2019hls}. Assuming all observations and analysis assumptions are correct and free of systematic effects, we are entering a regime where a direct detection of neutrino masses with particle detectors such as KATRIN could become a strong hint for nonstandard neutrino physics. At the same time, the next generation of cosmological surveys aim to improve their reach in neutrino mass sensitivity and to make the first detection~\cite{Abazajian:2013oma,Mishra-Sharma:2018ykh}.

To work toward these future goals, 
we present here a model-independent approach to investigate the possibility that neutrino masses change on cosmological timescales, reconstructing the neutrino mass as a function of redshift.

A similar methodology has been applied to reconstruct possible variations in other cosmological parameters, in particular dark energy parameters ~\cite{2010PhRvL.105x1302H,2011PhRvD..84h3501H,2012JCAP...06..036S,Seikel:2013fda,Vazquez:2012ce,Montiel:2014fpa,Hee:2015eba,Hee:2016nho,Poulin:2018zxs,Gerardi:2019obr,Joudaki:2017zhq,Keeley:2019esp,1840399,Bonilla:2021dql} and the Hubble parameter~\cite{2012PhRvD..85l3530S,2012PhRvD..86h3001S,Bernal:2016gxb,Poulin:2018zxs,Gomez-Valent:2018hwc,Renzi:2020fnx,Bernal:2021yli} as a function of redshift, the shape of the primordial power spectrum as a function of wave number~\cite{Bridle:2003sa,Guo:2011re,Hazra:2014jwa,Hunt:2013bha,Levy:2020emr}, and parameters describing deviations from general relativity as a function of redshift and wave number \cite{Garcia-Quintero:2020bac}. Exploring variations of cosmological parameters in time also gains in importance in light of tensions between parameters inferred from current high- and low-redshift data (see, e.g., Ref.~\cite{Verde:2019ivm}). 

In general, reconstructions fully accounting for degeneracies and the interplay between cosmological parameters, such as the dark matter density and dark energy parameters, are especially hard to achieve~\cite{Kunz:2007rk,Amendola:2012ky,Busti:2015aqa}. In the case of the neutrino mass, a model-independent reconstruction is particularly challenging. A comprehensive model for time-varying neutrino masses would require an interaction with other energy sectors, such as dark radiation or dark energy, to satisfy energy conservation laws. Additionally, including such interactions permits to fully capture and exploit the physics signatures of the model. While in the case of dark radiation this interaction is negligible for many cosmological scenarios (e.g., the dark radiation resulting from neutrino decays has negligible cosmological impact~\cite{Chacko:2019nej}), in the case of dark energy this interaction could alter $w_{\rm de}(z)$ and thus generate multiple signatures that allow us to place stronger limits on the model. 

In this paper, we take a conservative approach and use a generic function to model the neutrino mass sum. We also ignore potential additional constraining power coming from the inclusion of the coupling between the neutrino and the dark sector which is theory specific, and perform a model-independent analysis that only focuses on neutrinos without assuming any specific interactions with other dark sectors.
This yields a conservative approach in the modelling but not necessarily the most conservative neutrino mass constraints (see the discussion in Sec.~\ref{sec:conclusion}). The method used here expands on the work done in Ref.~\cite{Lorenz:2018fzb}, which assumed a specific time-varying neutrino mass model, and also spans models with neutrino decays for which we will set new limits. 

The paper is structured as follows. In Sec.~\ref{sec:theory} we explain the theoretical background of cosmological neutrino mass constraints and neutrino mass models. In Sec.~\ref{sec:methods} we present our methodology, in particular the different datasets and reconstruction methods. We present our results in Sec.~\ref{sec:results} and summarize and discuss them in Sec.~\ref{sec:conclusion}.

\section{Theoretical background}
\label{sec:theory} 
\subsection{Neutrino mass constraints from cosmological probes} \label{IIA}
\vspace{-0.3cm}
At the beginning of their cosmic journey, neutrinos are relativistic particles and behave as a radiation component in the early Universe. When their kinetic energy term drops below the mass term due to cooling in an expanding Universe, neutrinos start to behave as nonrelativistic, massive particles. The redshift of this transition is inversely proportional to the neutrino mass~\cite{Ichikawa:2004zi}: neutrinos with a smaller mass become nonrelativistic at a later time compared to neutrinos with a higher mass. Neutrino masses thus affect cosmological observations during different cosmic epochs, which in turn allows us to probe $\sum m_\nu$ at different redshifts. \\

\emph{The cosmic microwave background--} Constraining neutrino properties with the CMB has been a rich research area with extensive literature (see, e.g., Refs.~\cite{Archidiacono:2016lnv,lesgourgues_mangano_miele_pastor_2013,Vagnozzi:2017ovm}). To summarize, the key effects we look for in CMB probes are:
\indent (i) Effects on the background evolution of the Universe via changes in the angular diameter distance at recombination, $D_A(z_{rec})$ and the Hubble parameter~\cite{Archidiacono:2016lnv,lesgourgues_mangano_miele_pastor_2013}. In general, there is a strong degeneracy between neutrino parameters and the Hubble constant~\cite{Archidiacono:2016lnv,2012JCAP...04..027H,Sutherland:2018ghu,Moresco:2012by,2012MNRAS.425.1170H}. 

(ii) Effects on the evolution of perturbations. In particular, the integrated Sachs-Wolfe effect is affected, both at early times during radiation domination (eISW)~\cite{Archidiacono:2016lnv,lesgourgues_mangano_miele_pastor_2013}, as well at late times when dark energy starts to dominate the evolution of the Universe ($\ell$ISW)~\cite{Cabass:2015xfa}. The eISW depends on the relativistic degrees of freedom, $N_\mathrm{eff}$, as well on neutrino masses, but in a different way. Whereas $N_\mathrm{eff}$ mostly changes the amplitude of the first peak in the CMB anisotropy power spectrum, neutrino masses affect the amplitude of the eISW on a large range of multipoles depending on the neutrino mass and the corresponding free-streaming scale ~\cite{Archidiacono:2016lnv,lesgourgues_mangano_miele_pastor_2013,2014ApJ...782...74H}. The $\ell$ISW effect
dominates at $\ell<30$ and is therefore elusive because of cosmic variance in CMB data alone. However, when cross-correlated with galaxy number counts, this effect is a promising avenue for measuring neutrino masses~\cite{Lesgourgues:2007ix}. 

(iii) Effects on the matter distribution deflecting the CMB photons by gravitational lensing~\cite{Lewis:2006fu}. This will leave both an imprint on the CMB temperature and polarization anisotropies (in the high-$\ell$ region of the spectra), as well as generate a CMB lensing convergence signal, $C_\ell^{\phi\phi}$. The latter probe will capture the small-scale suppression of 
the matter power spectrum due to large neutrino thermal velocities and corresponding free-streaming out of density fluctuations~\cite{lesgourgues_mangano_miele_pastor_2013,Lesgourgues:2006nd}. Depending on whether the neutrino wavelength is above or below the free-streaming wavelength, neutrinos cluster as cold dark matter and baryons, or free-stream out of gravitational wells. This slows down the clustering of matter, leading to a suppression of the matter power spectrum on the corresponding scales which is more pronounced for larger neutrino masses~\cite{Hu:1997mj}.
The CMB damping tail is measured with high precision with current data~\cite{Henning:2017nuy,Aghanim:2019ame,Choi:2020ccd} and CMB lensing is now in a high signal-to-noise regime~\cite{Bianchini:2019vxp,Darwish:2020fwf}.

(iv) The optical depth to reionization $\tau$ is degenerate with the amplitude of scalar primordial fluctuations $A_s$. Since the amount of clustering of cosmic structures is tightly linked to $\sum m_\nu$, this in turn becomes a strong degeneracy between $\tau$ and $\sum m_\nu$. CMB polarization measurements of $\tau$ enable us to obtain a tighter constraint on $\sum m_\nu$~\cite{Allison:2015qca,Calabrese:2016eii}.\\ 

\emph{Baryonic acoustic oscillations--}
Before CMB decoupling, baryons and photons are tightly coupled to each other. At the time of recombination, the CMB photons decouple from the baryons, and the oscillations of the baryon-photon fluid are frozen in the CMB anisotropies. In addition to the CMB, these oscillations also leave a characteristic imprint in the large scale structure of the Universe, 
both transverse as well as along the line of sight~\cite{Eisenstein:2005su,Cole:2005sx}. In particular, the size of the BAO is known, and therefore BAO can be used as standard rulers to infer either $H(z)r_s(z^*)$ or $D_A/r_s(z^*)$, where $D_A(z)$ is the angular diameter distance (see, e.g., Refs.~\cite{2013PhR...530...87W,Bassett:2009mm} for reviews) and $r_s$ the sound horizon at decoupling $z^*$. 
These quantities are in particular sensitive to the matter density $\Omega_m h^2$. BAO measurements allow to constrain the energy contribution of massive neutrinos to the matter density. 
The BAO feature has been detected in the clustering of galaxies~\cite{Eisenstein:2005su,Cole:2005sx,Alam:2020sor}, as well as in the clustering of low-redshift quasars~\cite{deCarvalho:2017xye,Ata:2017dya,2021MNRAS.500.1201H}, and in the correlations of Lyman-$\alpha$ systems~\cite{McDonald:2006qs,Bautista:2017zgn} (see below). This latter probe provides an additional measurement to standard BAO data and extends BAO observations toward high redshifts.\\

\emph{Lyman-$\alpha$ forest--}
The absorption lines of neutral hydrogen in the intergalactic medium (IGM) in quasar spectra are sensitive to cosmological parameters, and probe cosmic structure formation at redshifts between $z\sim 2-6$. In particular, Lyman-$\alpha$ forest measurements characterize small structures on the scale of sub-Mpc to Mpc. This anchors the level of the matter power spectrum on scales between $k\sim 0.1-2$ and therefore probes the regime where the suppression of the matter power spectrum due to massive neutrinos is the most pronounced (see, e.g., Refs.~\cite{Seljak:2004xh,Seljak:2006bg,Palanque-Delabrouille:2014jca,Palanque-Delabrouille:2015pga,Palanque-Delabrouille:2019iyz}). The relevant summary statistic is the 1D flux power spectrum, which is related to the matter power spectrum using a nonlinear transformation~\cite{Croft_1998,Zhan:2005xs}. 
The level of the 1D flux power spectrum depends both on the amplitude of the linear matter power spectrum and on the sum of neutrino masses, resulting in a degeneracy
between these two parameters ~\cite{Pedersen:2019ieb}. This limits the ability of current Lyman-$\alpha$ forest measurements to constrain neutrino masses. The current upper bound from Lyman-$\alpha$ measurements alone is $\Sigma m_\nu<0.71$ eV  (95\% CL)~\cite{Palanque-Delabrouille:2019iyz}. 
Probing cosmological parameters with the Lyman-alpha forest is also challenging because of observational and astrophysical systematics, which need to be modeled in the analyses~\cite{Hui:1997dp,Walther:2020hxc}. \\

 \emph{Supernovae--}
Measurements of the luminosity distance $D_L(z)$ from supernovae explosions are not directly sensitive to neutrino properties. However, they can be used to constrain dark energy parameters, such as the dark energy density $\Omega_\Lambda$ and the dark energy equation of state $w_\mathrm{de}(z)$ (see, e.g., Ref.~\cite{2011ARNPS..61..251G} for a review), which helps significantly to break degeneracies with neutrino parameters. \\

Combining the different cosmological probes described above enhances significantly the constraining power coming from only one of the datasets (see, e.g., Refs.~\cite{Archidiacono:2016lnv,Abazajian:2013oma,Vagnozzi:2017ovm,Boyle:2018rva,Aghanim:2018eyx}). This is due to the fact that (i) different probes are capturing the physics of the Universe at different redshifts, and (ii) different probes depend on different parameter combinations, which follow different degeneracy directions. For example,  Ref.~\cite{Archidiacono:2016lnv} describes in detail how the combination of CMB and BAO helps to break the $H_0-\sum m_\nu$ degeneracy. 

\subsection{Neutrino mass models}\label{sec:models}
\vspace{-0.3cm}
Most of the possible neutrino mass models, in particular the ones arising from the famous seesaw mechanism~\cite{Minkowski1977,Gell-Mann1979,Yanagida1980,Mohapatra1980,Schechter1980,Schechter1982}, cannot be tested using cosmological data. From a cosmological perspective, more focus is then naturally placed on studying neutrino mass mechanisms that make cosmologically testable predictions, such as models providing neutrino mass variations on cosmologically interesting timescales.

As mentioned earlier, when allowing for the neutrino mass to vary as a function of redshift, an interaction with the dark energy or dark radiation sector is required to satisfy energy conservation laws. The most popular models that couple the neutrino and dark energy sectors are mass varying neutrino (MaVaN) scenarios (see, e.g., Ref.~\cite{Fardon:2003eh}). Here, the neutrinos couple to a light scalar field, which slowly rolls in a flat potential. Originally, this direct link between neutrino masses and dark energy was proposed to explain the similarity between the energy scales of neutrino masses and quintessence-like dark energy ($E\sim 10^{-3}$~eV). However, the coupling to a light scalar mediates an attractive force between the neutrinos and leads to bound state formation~\cite{Afshordi:2005ym}, such that the light scalar field can only explain dark energy under rather special circumstances (see, e.g., Refs.~\cite{Bjaelde:2007ki,Ayaita:2014una,Mandal:2019kkv}). For example, the Growing Neutrino Quintessence model studied in Ref.~\cite{Ayaita:2014una} yields time-dependent neutrino masses, which vanish in the early Universe and raise from $m_\nu =0$~eV at $a\lesssim 0.2$ to $m_\nu \sim 0.7$~eV at $a=1$.

Coupling the mass-varying neutrino sector to dark radiation is less trivial and usually occurs with simultaneously coupling both sectors to the dark energy sector. For example, it has been proposed that a simultaneous variation of neutrino masses, a light sterile neutrino fraction, and dark energy can be used to test certain aspects of the Weak Gravity Conjecture~\cite{DAmico:2018hgc}. 

Another gravitational avenue to couple neutrinos to dark energy and dark radiation is the gravitational neutrino mass model proposed in Ref.~\cite{Dvali:2016uhn}, which served as a motivation for the previous cosmological study in Ref.~\cite{Lorenz:2018fzb}. The key cosmological prediction of this model is that the cosmological neutrino mass parameter vanishes. Indeed, the model predicts massless neutrinos in the early Universe and the generation of neutrino masses in a late-time cosmological phase transition at \textit{$T\lesssim m_\nu$}. In the phase transition, neutrino masses are generated, followed by neutrino decay into the lightest mass eigenstate and rapid annihilation into massless Goldstone bosons (see Refs.~\cite{Dvali:2016uhn,Dvali:2016eay,Funcke:2019grs} for more details). Thus, the model predicts massless neutrinos in the early Universe and massless dark radiation in the late Universe. The intermediate regime with massive neutrinos, which exists directly after the phase transition, exists only for cosmologically negligible timescales. This implies that the cosmological neutrino mass parameter would be zero, as currently preferred by cosmological data~\cite{Aghanim:2018eyx}. This also implies that experiments aiming at a direct detection of the relic neutrino background, such as the proposed PTOLEMY experiment~\cite{Betti_2019}, would not be able to detect this background,\footnote{At first sight, one might expect that PTOLEMY could detect the massless Goldstone bosons, which are neutrino-composite bosons (similar to the light quark-composite mesons in QCD). However, the boson's energy is $E=T_\nu<m_\nu$, while the neutrino capture would release the other neutrino of the bound state, requiring an energy of at least $m_\nu$. This would violate energy conservation, unless one of the neutrinos is almost massless. We thank Pedro Machado for bringing up this argument.} unless we allow for substantial neutrino asymmetries (see below).

If the model in Ref.~\cite{Dvali:2016uhn} is extended by allowing for neutrino asymmetries (i.e., more neutrinos than antineutrinos or vice versa), not all neutrinos would find an antineutrino partner to annihilate, leaving behind a fraction of relic neutrinos with a nonzero cosmological mass parameter. As Ref.~\cite{Lorenz:2018fzb} demonstrated, even in this case, the cosmological neutrino mass bound would be substantially weakened to $\sum m_\nu <4.8$~eV (95\% CL). However, we note that this weakened bound was obtained with \textit{Planck} 2015
data and is expected to become more stringent when including the
\textit{Planck} 2018 dataset. This is because the $\Lambda$CDM constraint on $\sum m_\nu$ strengthened by a factor of $\mathcal{O}(2)$ from $\sum m_\nu=  0.21$~eV  (95\%  CL, \textit{Planck}  2015  without  polarization~\cite{Ade:2015xua}) to $\sum m_\nu= 0.12$~eV (95\%  CL,  \textit{Planck}  2018~\cite{Aghanim:2018eyx}). 

Finally, neutrinos can couple to the dark radiation sector if they are unstable, with lifetimes shorter than the age of the Universe. In particular, 2-body decays of neutrinos into BSM particle species, such as massless or very light sterile neutrinos and Goldstone bosons, have been proposed to relax cosmological neutrino mass bounds. Early cosmological studies based on \textit{Planck} 2015 data implied a relaxation of the cosmological neutrino mass bound up to $\sum m_\nu \lesssim 0.9$~eV~\cite{Chacko:2019nej,Chacko:2020hmh,Escudero:2020ped}. As above, this bound is expected to become tighter when including \textit{Planck} 2018 data (see Sec.~\ref{sec:decay}). Thus, neutrino decay can only slightly alleviate cosmological bounds, similar to other extensions of the $\Lambda$CDM model, such as dynamical dark energy. However, such models leave the intriguing possibility of observing less (or even completely vanishing) neutrino mass in the late Universe, similar to the model proposed in Ref.~\cite{Dvali:2016uhn}, and thus make cosmological observations crucial for determining the neutrino mass origin and beyond-SM neutrino interactions.

These and other theoretical motivations to study neutrino mass variations on cosmologically interesting timescales, which typically yield nontrivial interactions with other cosmological sectors such as dark energy and dark radiation, are the main rationale for the work that we present here. 

\section{Methodology} 
\label{sec:methods}

\begin{table*}[t]
\begin{tabular}{l|l}
Dataset & Redshift range\\
\hline
\hline 
Planck 2018 CMB TTTEEE~\cite{Aghanim:2018eyx,Aghanim:2019ame} & mostly $z=1100$\\
Planck 2018 CMB  lowl~\cite{Aghanim:2018eyx,Aghanim:2019ame} & mostly $z=1100$\\
Planck 2018 CMB  lowE~\cite{Aghanim:2018eyx,Aghanim:2019ame} & mostly $z=8$ and $z=1100$\\
Planck 2018 CMB lensing~\cite{Aghanim:2018eyx,Aghanim:2018oex} & $0\leq z\leq 1100$\\ 
\hline 
BAO (6dF)~\cite{2011MNRAS.416.3017B} & $z=0.106$\\ 
BAO (SDSS DR7 BOSS MGS)~\cite{Ross:2014qpa} & $z=0.15$ \\
BAO (SDSS DR12 BOSS)~\cite{Alam:2016hwk} & $z=0.38,0.51, 0.61$\\ 
BAO (SDSS DR14 eBOSS quasars)\cite{Ata:2017dya} & $z=1.52$\\ 
BAO (SDSS DR14 eBOSS Ly-$\alpha$)~\cite{Blomqvist:2019rah} & $z=2.34$ \\
BAO (SDSS DR14 eBOSS cross  Ly-$\alpha$-QSO)~\cite{Blomqvist:2019rah} & $z=2.35$\\
\hline 
SN (Pantheon)~\cite{Scolnic_2018} & $0.01<z<2.3$\\ 
\hline
\hline
\end{tabular} 
\caption{\label{tab:datasets} List of datasets used in the analyses performed in this work and their corresponding redshift range.}
\end{table*}

\subsection{Data}
\label{sec:data}
\vspace{-0.3cm}
Motivated by the expected contribution highlighted in Sec.~\ref{IIA}, we include the following datasets in our analysis: 
\begin{enumerate}[label={\arabic*.)}]
	\item\textbf{CMB and CMB lensing:} We use CMB temperature, polarization and lensing data from the \textit{Planck} 2018 data release~\cite{Aghanim:2018eyx}.
	
	\item\textbf{BAO:} We use BAO from 6dF~\cite{2011MNRAS.416.3017B}, the Sloan Digital Sky Survey (SDSS) DR7 Main Galaxy Sample (MGS)~\cite{Ross:2014qpa} and the SDSS Baryon Oscillation Spectroscopic Survey (BOSS) twelfth data release (DR12)~\cite{Alam:2016hwk}.  The mild discrepancies seen between \textit{Planck} and Lyman-alpha BAO data have decreased in recent releases of the Lyman-alpha BAO from SDSS DR14 eBOSS~\cite{Blomqvist:2019rah,Schoneberg:2019wmt}. Therefore we include also this additional BAO dataset which provide two data points at $z\sim 1.5$ and $z\sim2.3$. We also include the BAO measurement from quasars from the 14th data release of the extended BOSS (eBOSS) quasar sample \cite{Ata:2017dya}, giving us another data point at $z\sim 1.5$. 
	
	\item\textbf{Supernovae:} We additionally add type IA supernova data from from the Pantheon Supernovae Sample~\cite{Scolnic_2018} in order to break the degeneracies between dark energy and neutrino masses. 
	
\end{enumerate}

We demonstrate the impact of these specific datasets later in Sec.~\ref{sec:results}, and summarize their respective redshift range in Tab.~\ref{tab:datasets}.

\subsection{Reconstruction method}
\label{sec:rec}
\vspace{-0.3cm}
We modify the publicly available Einstein-Boltzmann code \texttt{CAMB}~\cite{Lewis:1999bs} and the corresponding Monte-Carlo Markov chain package \texttt{CosmoMC}~\cite{Lewis:2002ah} in order to implement different reconstruction methods for redshift-dependent neutrino masses. 

\subsubsection{Binned reconstruction}
\vspace{-0.3cm}
As a first attempt, we parametrize the neutrino mass with a step function for which it is assumed that the neutrino mass has a constant positive amplitude in each redshift bin. In the statistics literature this is  called a regressogram \cite{Wasserman06}. To maximally exploit our datasets we use a parametrization with six redshift intervals: 
\begin{equation} 
\sum m_\nu(z)=\begin{cases}
\sum m_{\nu,0} &\text{($0\leq z < z_1$)} \\
\sum m_{\nu,1} &\text{($z_1\leq z < z_2$)} \\
\sum m_{\nu,2} &\text{($z_2\leq z < z_3$)} \\
\sum m_{\nu,3} &\text{($z_3\leq z < z_4$)} \\
\sum m_{\nu,4} &\text{($z_4\leq z < z_5$)} \\
\sum m_{\nu,5} &\text{($z\geq z_5$)} . 
\end{cases} 
\end{equation}
We choose the edges of the redshift bins to pinpoint specific transitions in the composition of the Universe and to highlight the impact of using different cosmological probes. We set $z_1=0.5$, $z_2=3$, $z_3=10$, $z_4=100$ and $z_5=1100$. The first bin explores $\sum m_\nu$ during dark energy domination, the second bin spans the BAO interval, the third bin includes most of the remaining information expected in CMB lensing~\cite{Carbone:2007yy}, the fourth bin covers intermediate redshifts, and the second to the last bin stretches out to the time of CMB decoupling. The last bin for $z>1100$ captures all the integrated information of the pre-recombination Universe. This choice of bins will also allow us to set constraints on neutrino decay models~\cite{Escudero2019,Chacko:2019nej,Chacko:2020hmh,Escudero:2020ped}, as described in Sec.~\ref{sec:decay}. The neutrino masses in the individual bins are five additional parameters in the model, compared to the standard $\Lambda$CDM model with massive neutrinos. The discontinuity in the binned $\sum m_\nu(z)$ leads only to small discontinuities in related quantities, such as the Hubble rate $H(z)$, which however cause no numerical instabilities in the analysis.
A similar parametrization has recently been used for the reconstruction of the dark energy equation of state $w_\mathrm{de}(z)$ from gamma-ray bursts~\cite{2020arXiv201203392M}. 

\subsubsection{Spline priors} 
\vspace{-0.3cm}
To perform a smoother fit 
and to potentially identify features in the neutrino mass sum that are hidden in the binned reconstruction, we also consider Bayesian regression splines with variable knot points~\cite{DimatteoGenoveseKass01}.

The knots correspond to the bin margins of the regressogram prior function described above, and model change points where the trajectory of the neutrino mass might change its slope. Previous literature in cosmology 
has often used splines with fixed knot positions (see, e.g.~\cite{Guo:2011re,Bernal:2016gxb,Poulin:2018zxs}). This, however, requires to choose the positions of the knots in advance and can therefore significantly influence or bias the result of the reconstruction.
Here, we estimate the position of the knots from the data and include them as free parameters in the analysis. 
Using variable knot positions yields a more flexible fit than in the case with fixed knots and allows the fit to adapt to underlying features of the model. This methodology has previously successfully been applied several times in cosmology~\cite{2012JCAP...06..006V,Vazquez:2012ce,Millea:2018bko}, most recently in Ref.~\cite{Bernal:2021yli}.

Compared to binned priors, the resulting reconstruction is smooth and not piecewise constant anymore. 
We model two knots, $z_1$ and $z_2$, and linearly interpolate between $\sum m_{\nu,z_0}$, $\sum m_{\nu,z_1}$, $\sum m_{\nu,z_2}$ and $\sum m_{\nu,z_3}$. For our analysis, we choose $z_0=0$ and $z_3=1100$. In that case, the neutrino mass at redshift $z$ is given by 
\begin{align} \label{eq:splines}
&\sum m_\nu(z)\\
=&\begin{cases}
\sum m_{\nu,z_0}+(\sum m_{\nu,z_1}-\sum m_{\nu,z_0})\frac{z}{z_1}&\text{($z_0\leq z < z_1$)} \notag \\
\sum m_{\nu,z_1}+(\sum m_{\nu,z_2}-\sum m_{\nu,z_1})\frac{z-z_1}{z_2-z_1} &\text{($z_1\leq z < z_2$)} \notag \\
\sum m_{\nu,z_2}+(\sum m_{\nu,z_3}-\sum m_{\nu,z_2})\frac{z-z_2}{z_3-z_2} &\text{($z_2\leq z < z_3$)} \notag \\
\sum m_{\nu,z_3} &\text{($z\geq z_3$)}\notag  .
\end{cases} 
\end{align}

We then choose a uniform prior on the logarithm of the corresponding redshift with $-1\leq \log_{10}(z_i)_{i=1,2}\leq 3.041$, covering the entire redshift region until recombination and focusing on low redshifts where most of the data is located. We further need to impose that $z_2$ has to be equal or larger than $z_1$, in accordance to the definition above in Eq.~\eqref{eq:splines}. In addition, we choose $z_{1,2}>0.1$, corresponding to the lowest redshift of the BAO dataset (see Table~\ref{tab:datasets}). 

With this parametrization, we have five additional free parameters compared to the standard $\Lambda$CDM with massive neutrinos, three for $\sum m_\nu$ at $z_1$, $z_2$ and $z_3$, and two for the positions of $z_1$ and $z_2$. In order to obtain pointwise credible bands for $\sum m_\nu(z)$, we compute the neutrino mass sum for two hundred points in $z$ as derived parameters with \texttt{GetDist}~\cite{Lewis:2019xzd} to sample well the entire redshift range. We then compute pointwise the mean and credible intervals for each point in $z$.

\section{Constraints from current data}
\label{sec:results} 
\vspace{-0.3cm}
For our neutrino mass reconstruction, we run MCMC chains with our modified version of \texttt{CosmoMC}. Since both our methods only touch the neutrino mass modeling, we vary the standard $\Lambda$CDM parameters alongside with the new parameters of the reconstruction: the cold dark matter density $\Omega_{c} h^2$, the baryon density $\Omega_{b} h^2$, the scalar spectral index $n_s$, the amplitude of primordial fluctuations $A_s$ and the optical depth to reionization $\tau$. In addition, we have six additional parameters for the reconstruction with fixed bins ($\sum m_{\nu,0}$, $\sum m_{\nu,1}$, $\sum m_{\nu,2}$, $\sum m_{\nu,3}$, $\sum m_{\nu,4}$ and $\sum m_{\nu,5}$), and six for the reconstruction with linear splines and variable knots ($\sum m_{\nu,z_0}$, $\sum m_{\nu,z_1}$, $\sum m_{\nu,z_2}$, $\sum m_{\nu,z_3}$ and the position of the knots $\log_{10}(z_1)$ and $\log_{10}(z_2)$). We choose a uniform prior between [0:5] for $\sum m_\nu$ in the individual redshift bins (similarly to the \textit{Planck} 2018 analysis~\cite{Aghanim:2018eyx}), and we do not assume a specific correlation between the values of $\sum m_\nu$ in the different bins, leaving them to vary independently. In addition, we assume a normal neutrino mass hierarchy in line with current cosmological constraints~\cite{Aghanim:2018eyx,Gariazzo:2018pei,RoyChoudhury:2019hls}.
The choice of the mass hierarchy should not significantly affect our final results presented in Sec.~\ref{sec:results}, as current cosmological data cannot (yet) distinguish between the different neutrino mass hierarchies~\cite{Jimenez:2010ev,Gerbino:2016ehw,Vagnozzi:2017ovm,Lattanzi:2017ubx,Gariazzo:2018pei,deSalas:2018bym,RoyChoudhury:2019hls,Mahony:2019fyb,Archidiacono:2020dvx,Xu:2020fyg,Stocker:2020nsx,Hergt:2021qlh}.  

\subsection{Binned reconstruction}
\begin{figure*}[ht!]
\centering
    \includegraphics[width=0.78\columnwidth]{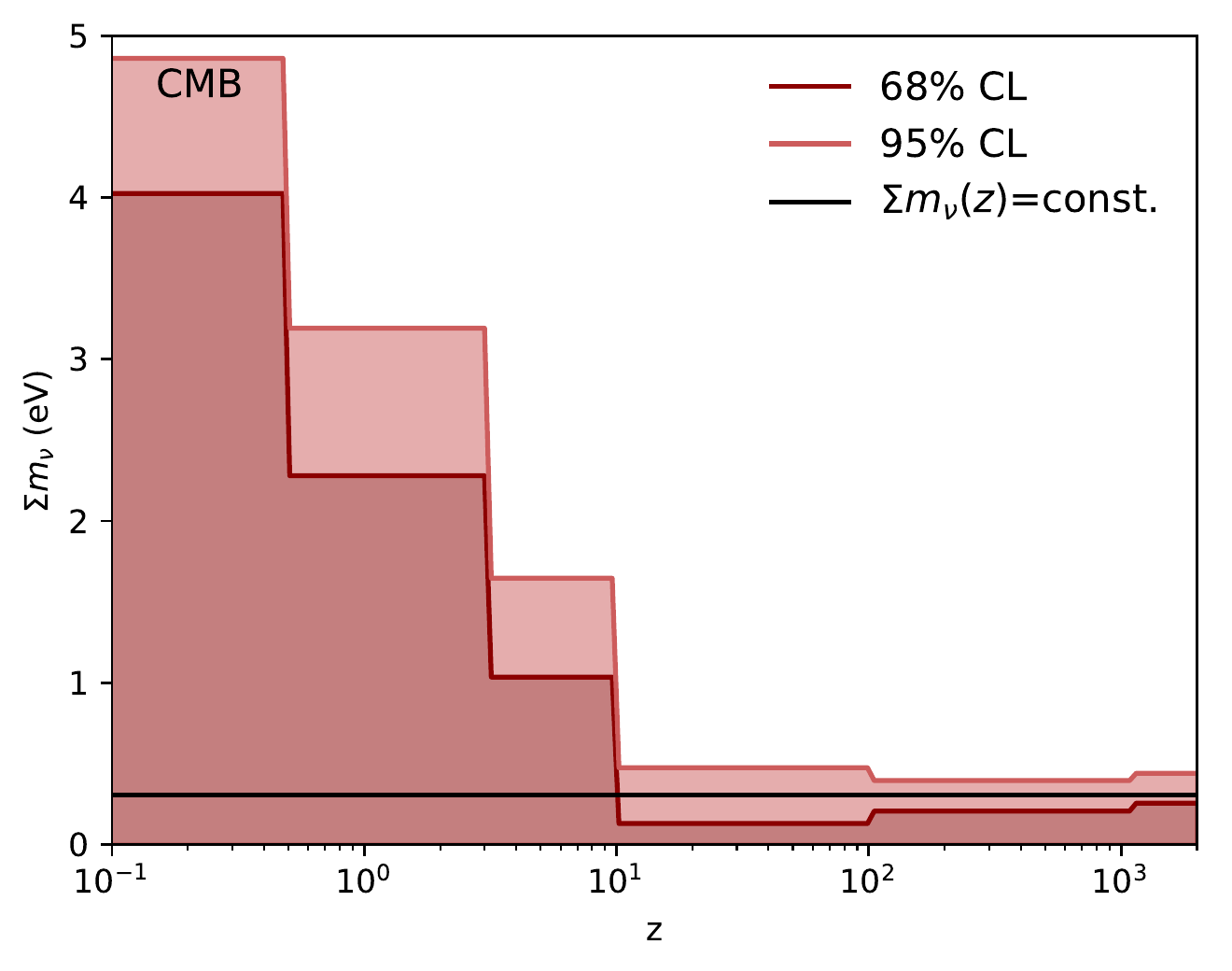}
    \includegraphics[width=0.78\columnwidth]{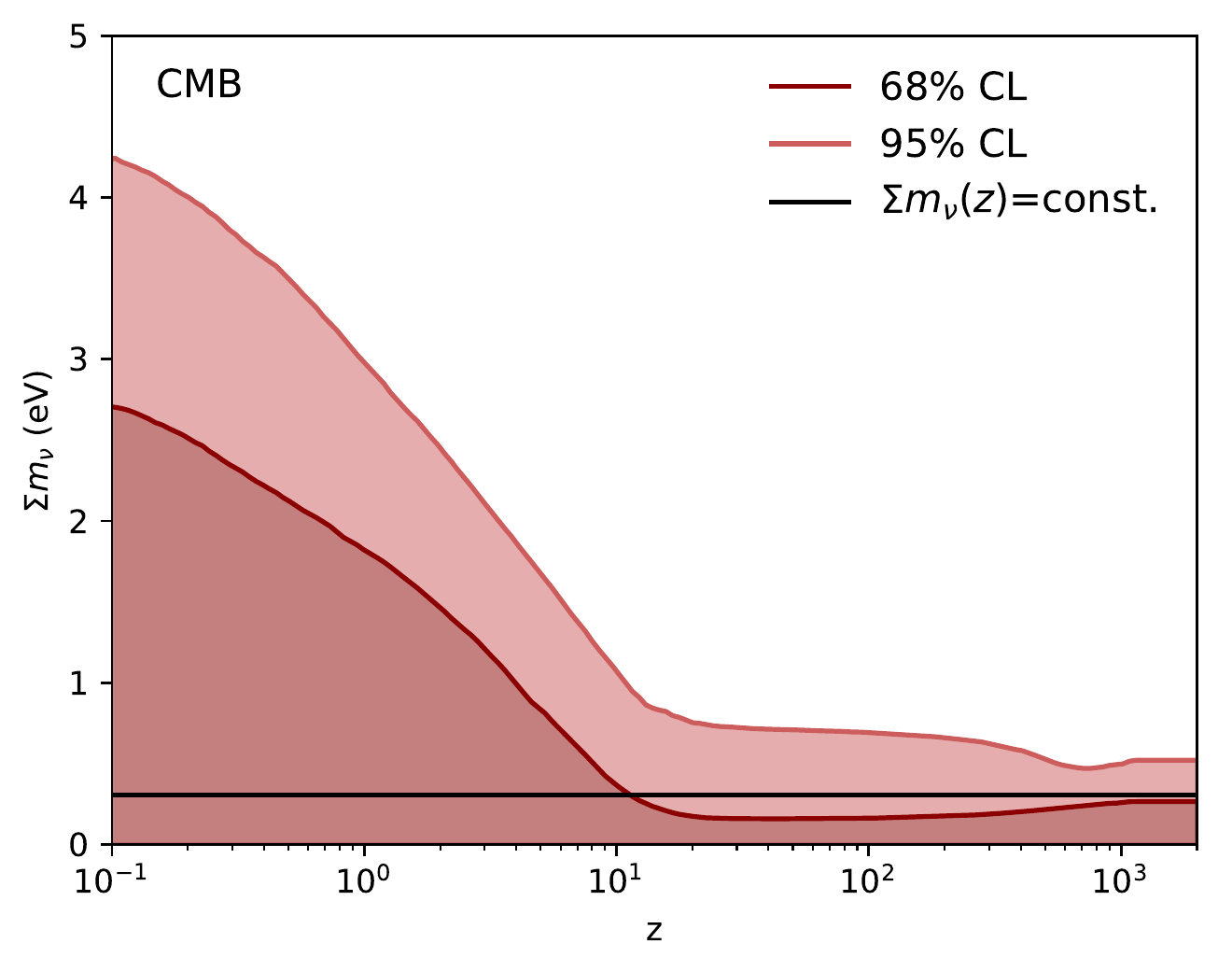}
    \includegraphics[width=0.78\columnwidth]{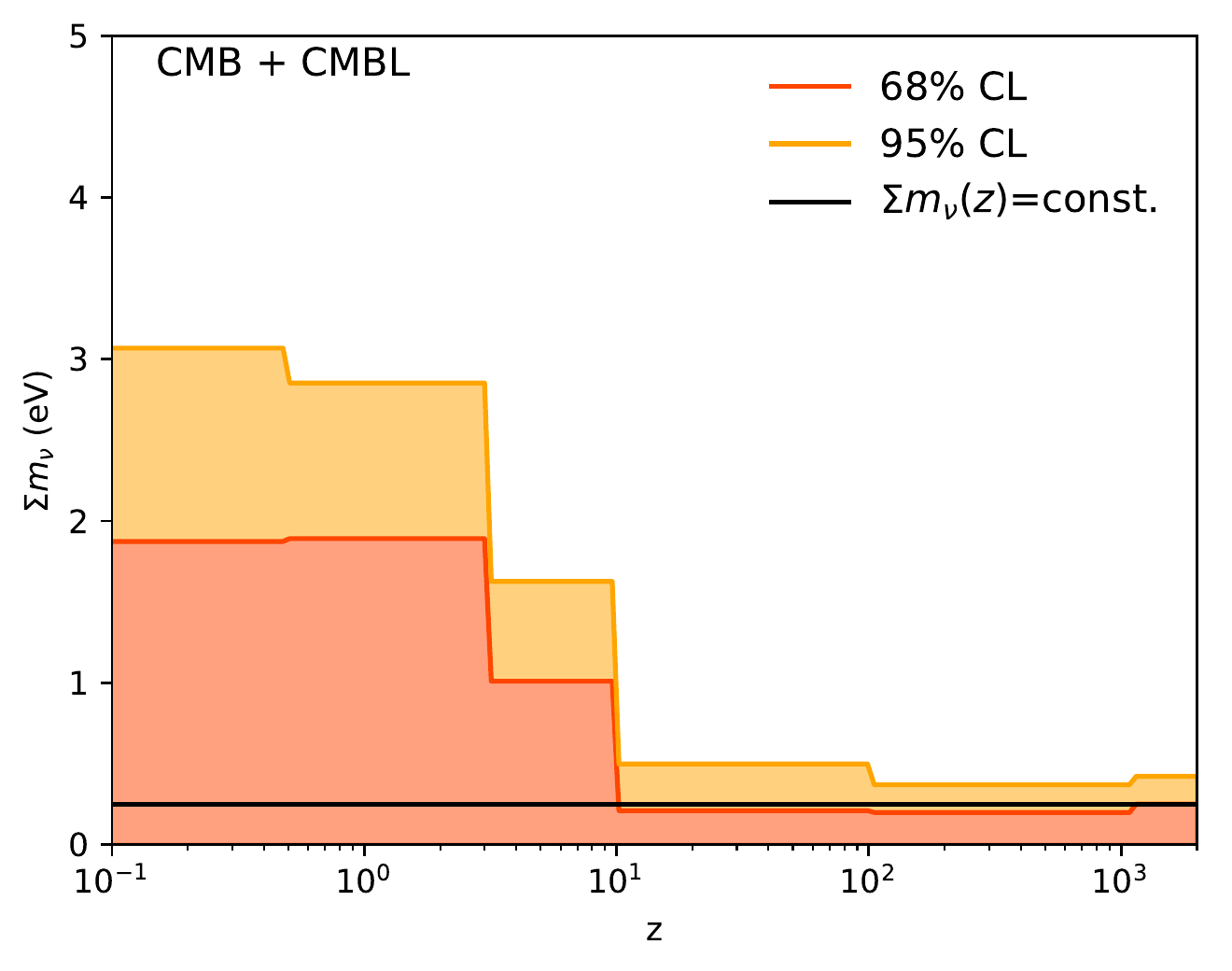}
    \includegraphics[width=0.78\columnwidth]{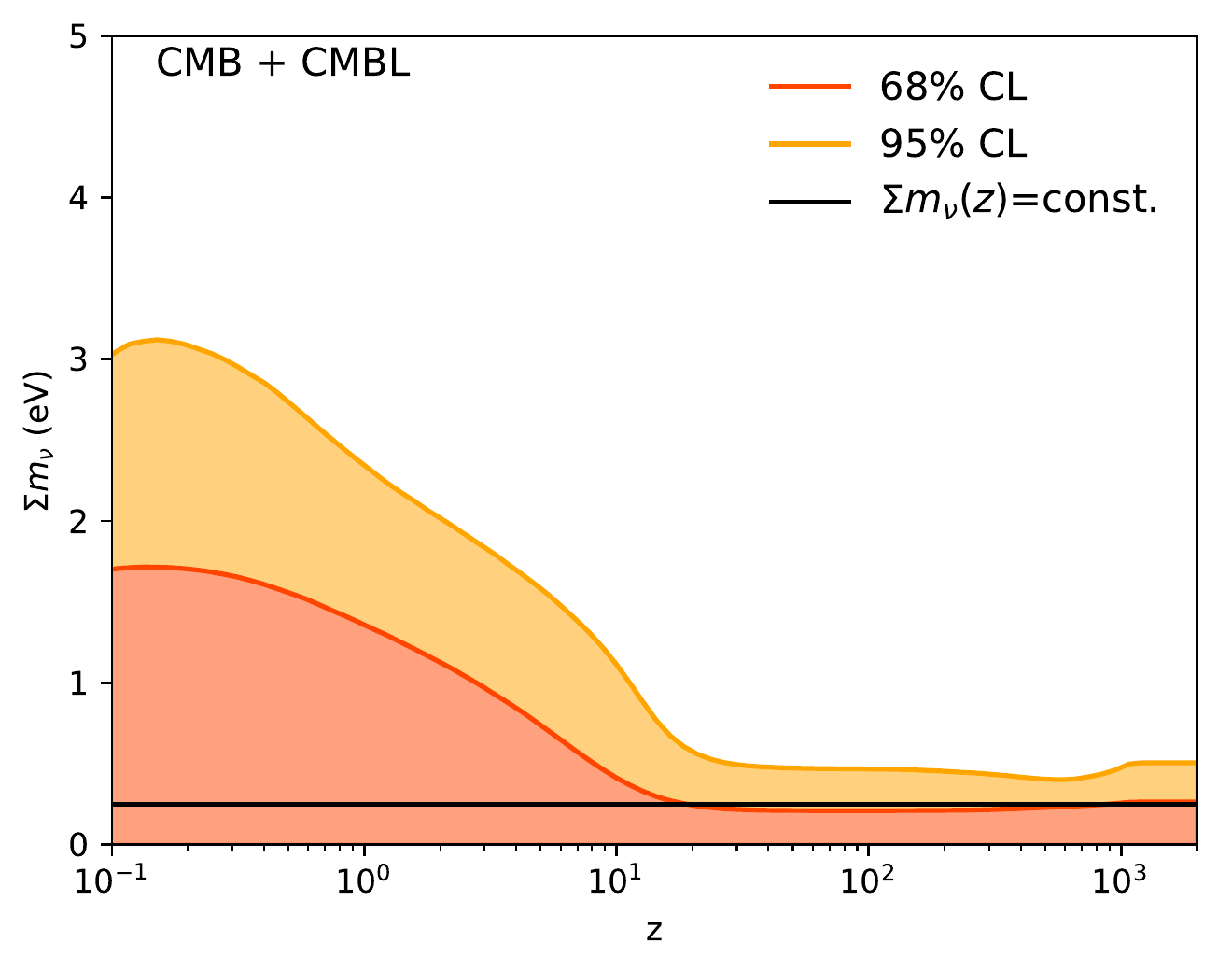}
    \includegraphics[width=0.78\columnwidth]{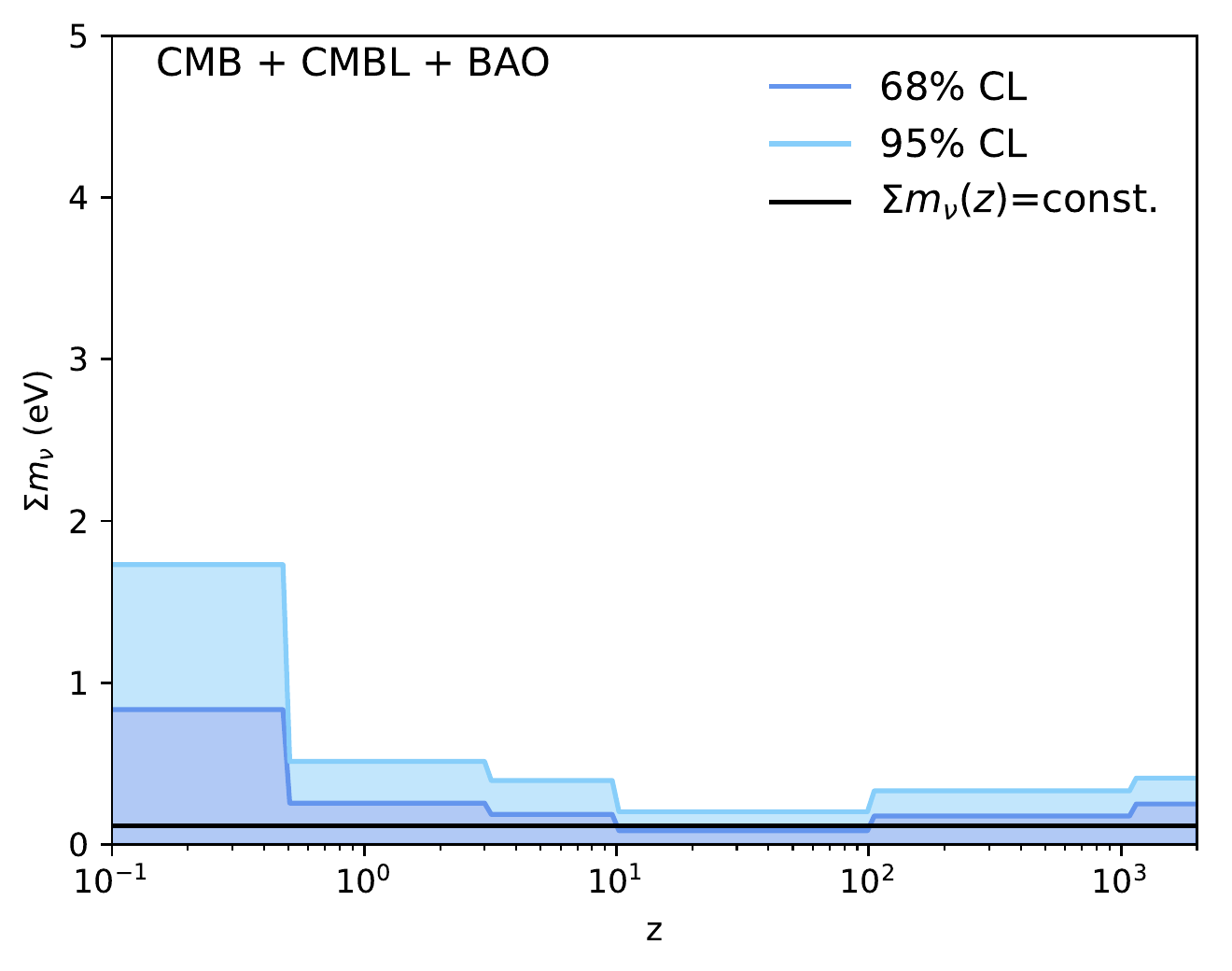}
    \includegraphics[width=0.78\columnwidth]{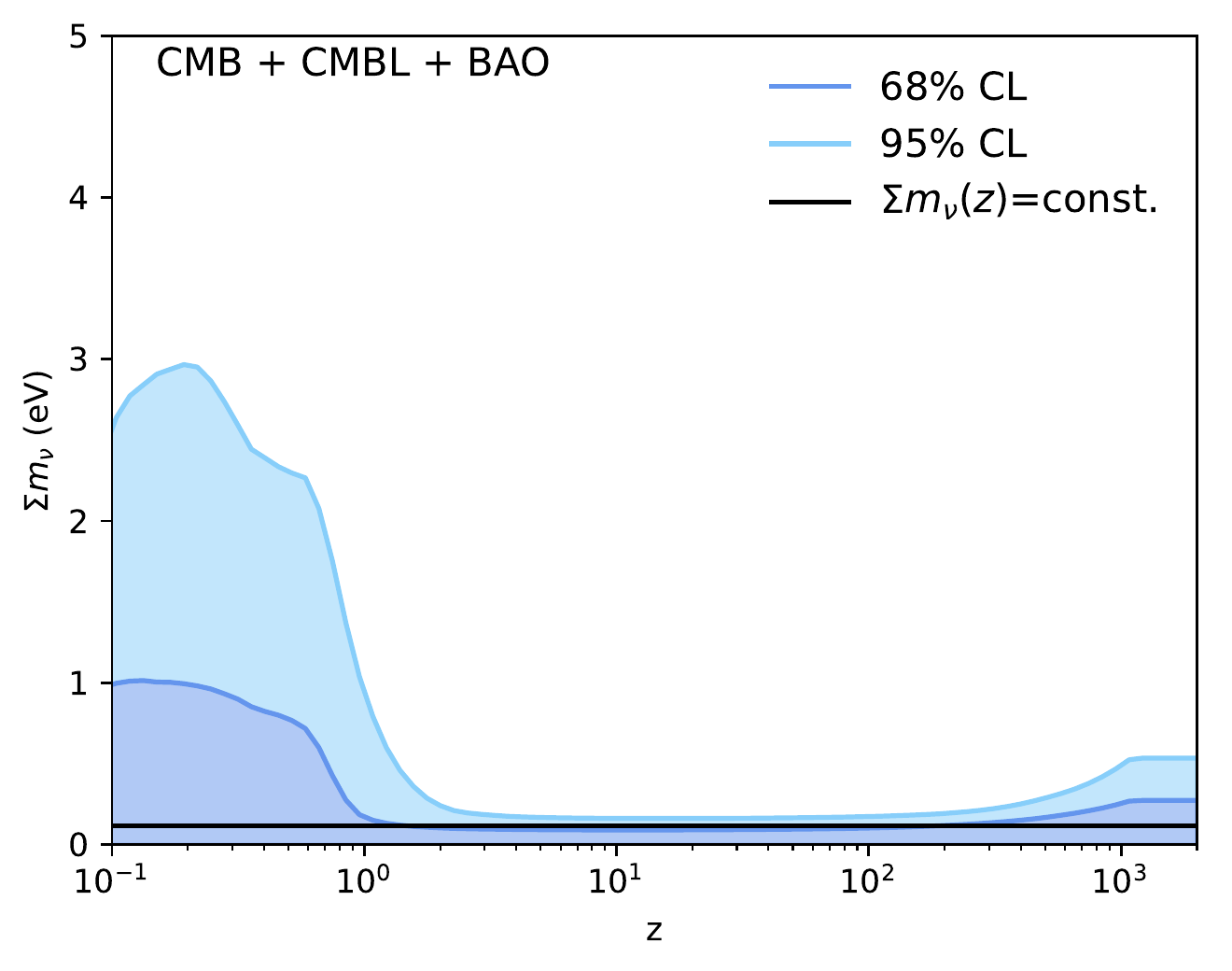}
    \includegraphics[width=0.78\columnwidth]{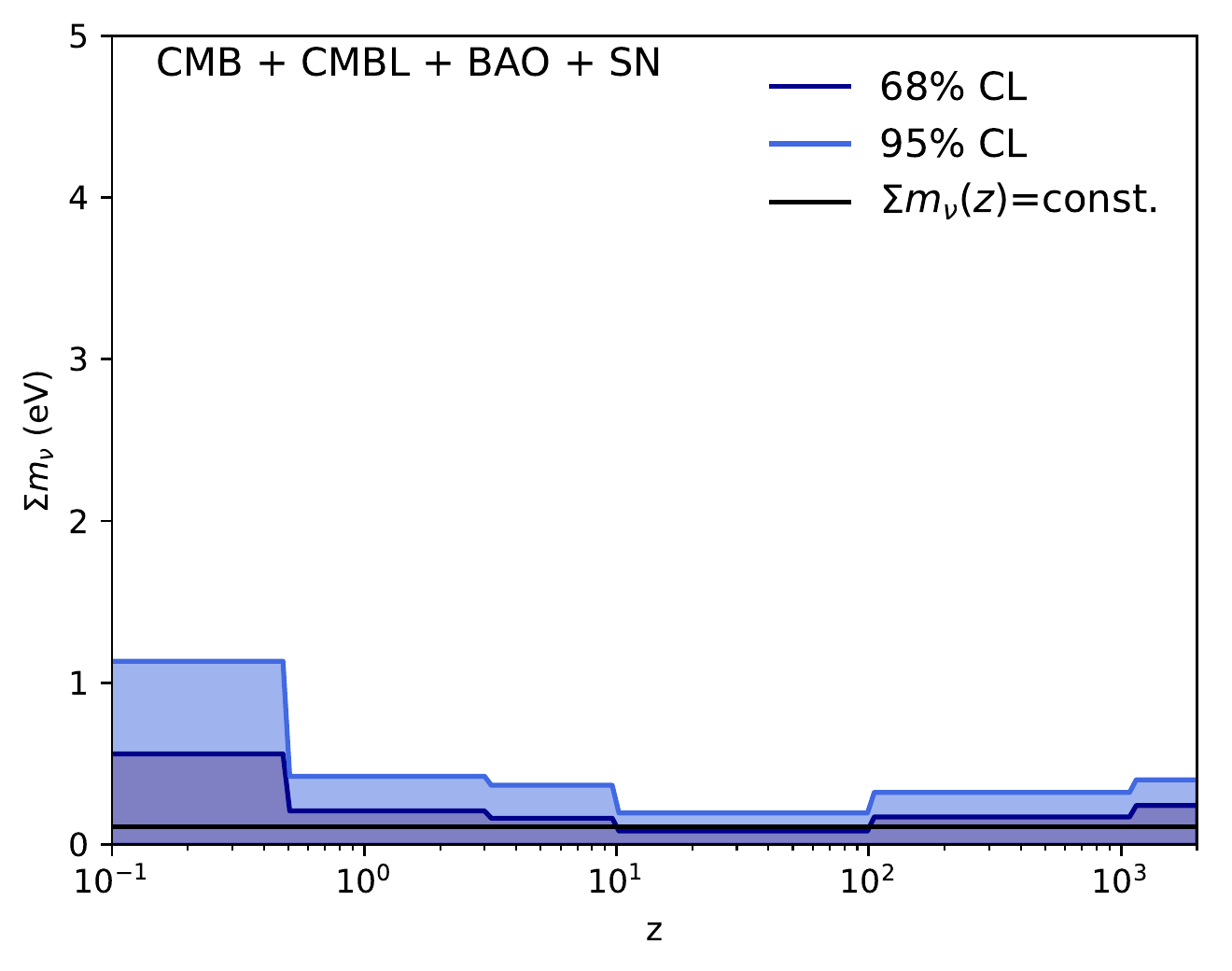}
    \includegraphics[width=0.78\columnwidth]{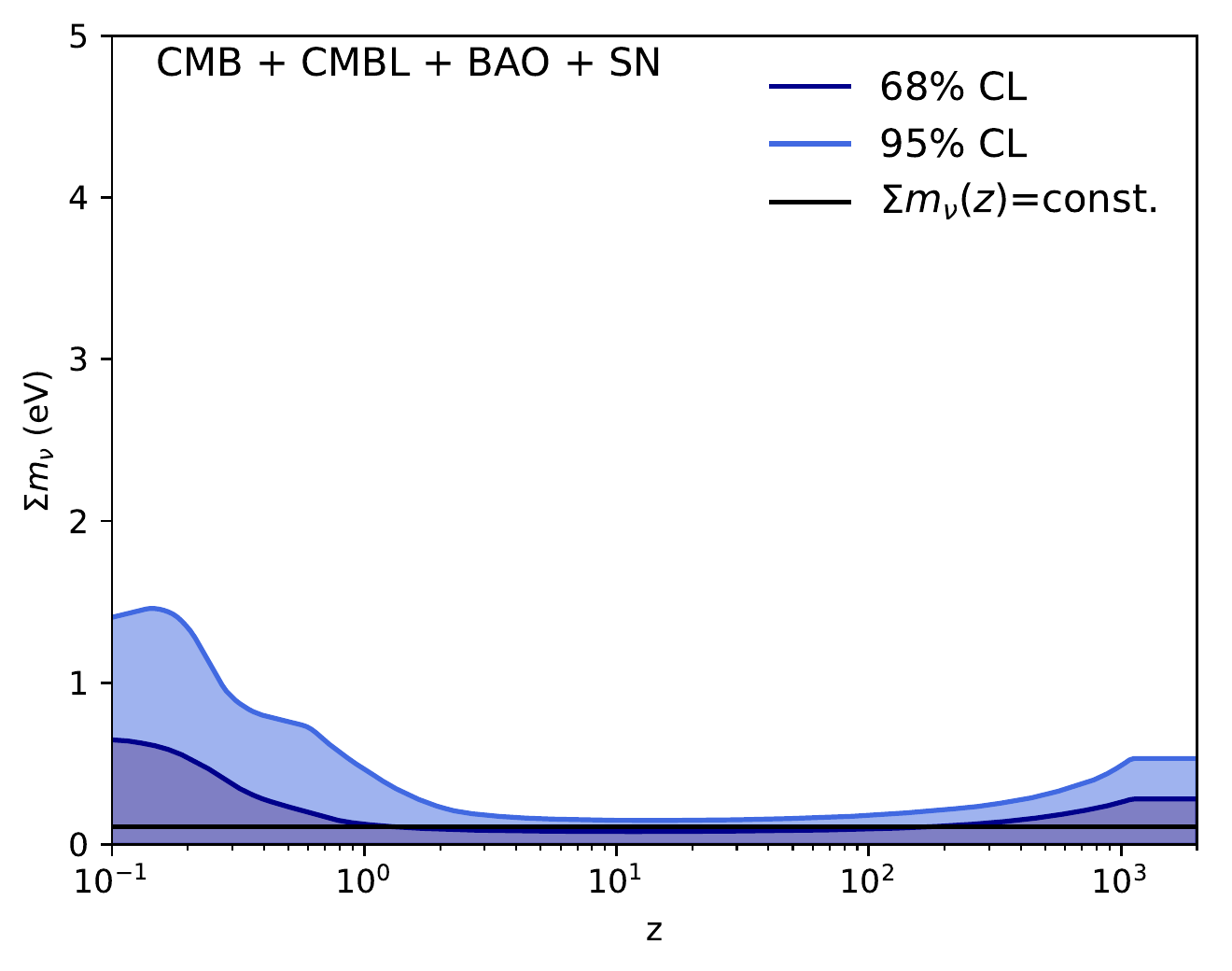}
    \caption{Reconstructed upper bounds of the neutrino mass as function of redshift. The panels from top to bottom show incremental addition of data starting from \textit{Planck} 2018 CMB temperature and polarization, and then  subsequently adding \textit{Planck} CMB lensing, BAO from BOSS DR12, 6dF and MGS, eBOSS DR14 quasars and Lyman-alpha, and SN from Pantheon 18. The left panels show the results obtained with the binned parameterization and the right panels the limits from the reconstruction with linear splines and variable knots. In all cases, we also report with the black solid line the 95\% CL constraint from the same data combination for $\sum m_\nu(z)={\rm const.}$ (i.e., the standard single parameter extension of $\Lambda$CDM for massive neutrinos).} 
    \label{fig:mnuz_data}
\end{figure*}

\vspace{-0.3cm}
We show the results of the binned neutrino mass reconstruction in Fig.~\ref{fig:mnuz_data}, and in Table~\ref{tab:results}. We plot both the 95\% and the 68\%~CL upper limits for each mass parameter, as well as the 95\% limit for $\sum m_\nu(z)={\rm const. }$ (i.e., the standard single parameter extension of $\Lambda$CDM for massive neutrinos) obtained from the same data combinations. We always plot the 95\% CL and 68\% CL one-tail upper limits, e.g., we have removed the 32\% and 5\% highest samples for these limits. For consistency we choose the same type of limits for all the individual redshift bins, and can therefore compare better the results for the individual bins and data combinations. For the full data combination including CMB, CMB lensing, BAO and SN, we find 
\begin{align*}
    \sum m_\nu(0\leq z<0.5)&<1.13 \text{  eV}\\
    \sum m_\nu(0.5\leq z<3) &<0.42 
    \text{  eV}\\
    \sum m_\nu(3\leq z<10) &<0.37 
    \text{  eV}\\
    \sum m_\nu(10\leq z<100) &<0.19 
    \text{  eV}\\
    \sum m_\nu(100\leq z<1100) &<0.32
    \text{  eV}\\
    \sum m_\nu(z\geq 1100)&<0.40 \text{ eV}\hspace{1em}\text{(95\% CL)}.
\end{align*}

These results are consistent with neutrino masses constant in time. In addition, we observe that the neutrino mass bound becomes less stringent at low redshifts ($z\leq 3$), and at very high redshifts ($z\geq 1100$). The large neutrino mass bound at low redshifts could arise due to the well-known degeneracy between the  neutrino mass sum and the dark energy density (see below). It could also point at a very late generation of neutrino masses in the Universe, 
and would therefore be consistent with the trend seen in Ref.~\cite{Lorenz:2018fzb}. In the Standard Model of Cosmology, the neutrino masses are generated during the electroweak or earlier phase transitions. \\

Our plots also allow us to note the contribution to the constraint from different probes. \\

\begin{figure*}[ht!]
    \centering
    \includegraphics[width=\textwidth]{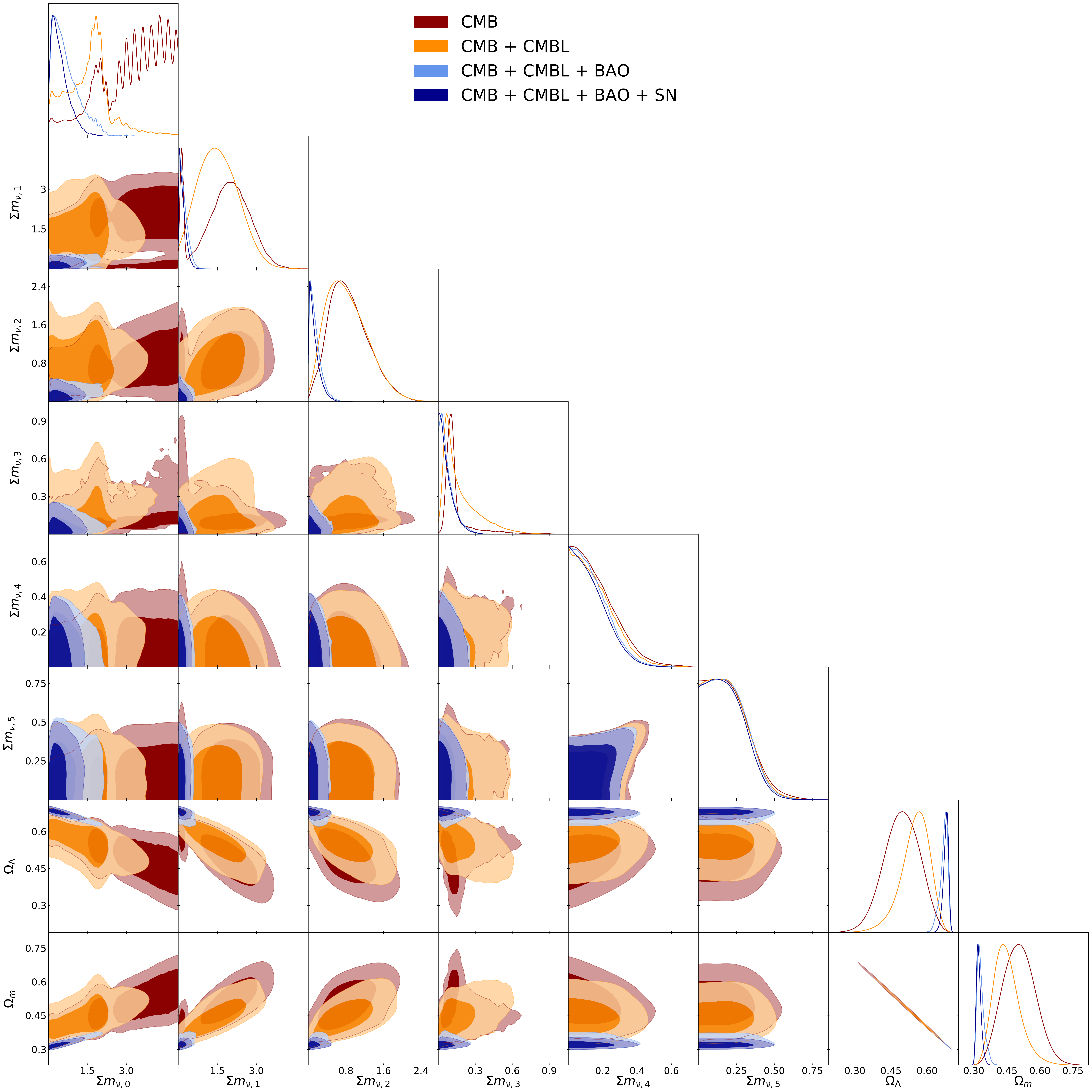}
    \caption{2-dimentional contours showing the 68\% and 95\% CL and 1-dimensional posteriors for the neutrino mass parameters in the individual bins, $\Omega_\Lambda$ and $\Omega_m$. Different colors show different data combinations, as in Fig.~\ref{fig:mnuz_data}.}
    \label{fig:mnu_lambda}
\end{figure*}

\noindent \emph{Top row:} In the top row of Fig.~\ref{fig:mnuz_data}, we plot the $z$-dependent constraints on the neutrino mass sum from CMB anisotropy data alone. Because the binned reconstruction introduces more free parameters, the constraints in all bins are consistent but systematically larger than the standard neutrino mass bound from the latest \textit{Planck} release~\cite{Aghanim:2018eyx,10.1093/ptep/ptaa104}\footnote{Note that the $\sum m_\nu(z)=$~const.\ line should be compared to the 95\% shaded region.}.  The CMB anisotropy data put a strong constraint on $\sum m_\nu$ at high redshifts, but only a weak constraint at low redshifts. We also note that the neutrino mass sum in the lowest redshift bin is unconstrained. In this bin the CMB data alone provides insufficient constraining power.
As described earlier, at low redshifts, the CMB anisotropy data would mostly be sensitive to neutrino masses via the late integrated Sachs-Wolfe effect at low multipoles $\ell$, which is cosmic variance limited and subject to significant parameter degeneracies. In the lowest redshift bin ($0<z<0.5$), the CMB has only little constraining power, and the prior (with prior mean $\sum m_\nu= 2.5$ eV) dominates the result of the reconstruction. At intermediate redshifts, the data strongly constrain the neutrino mass through the lensing of the CMB anisotropies which will be picked up here as a smoothing of the temperature and polarization data, and the impact of neutrinos on the background evolution (see the discussion in Sec.~\ref{sec:theory}). At high redshifts close to recombination, we observe a small increase in the uncertainty of $\sum m_\nu$. This might be due to the fact that at recombination the constraint on $\sum m_\nu$ comes mostly from the early integrated Sachs-Wolfe effect, the CMB damping tail, and background effects, while all other effects from neutrino masses only become relevant at intermediate to low redshifts. \\ \\
\noindent\emph{Second row:} In the second row in Fig.~\ref{fig:mnuz_data}, we plot the constraints on $\sum m_\nu(z)$ from CMB anisotropies plus the CMB lensing reconstruction. The CMB lensing kernel peaks around $z\sim 2$ (see, e.g., Refs.~\cite{Hassani:2015zat,Manzotti:2017oby}) and carries constraining power on the bin between $0.5\leq z<3$. Other constraints are also improved by an indirect effect: CMB lensing is also sensitive to the total matter density and the dark energy density~\cite{2011PhRvL.107b1302S}, and therefore helps to constrain $\sum m_\nu(0<z<0.5)$ by partially breaking the tridimensional degeneracy between $\Omega_\Lambda$, $\Omega_m$, and $\sum m_\nu$. This can also be seen in Fig.~\ref{fig:mnu_lambda}, where we show the results for $\sum m_\nu$ in the different bins, as well as its correlation with $\Omega_\Lambda$ and $\Omega_m$. \\ \\
\noindent\emph{Third row:} The BAO data points used in this work are distributed over the whole redshift range between $z=0.1-2.35$. This mostly improves the constraints in the first two neutrino mass bins. In addition, the inclusion of BAO reduces significantly the degeneracy between $\Sigma m_\nu$ and $\Omega_m$, and therefore also improves the overall constraints in all redshift bins. \\ \\
\noindent\emph{Bottom row:} We include SN data from the Pantheon 18 data set to constrain further the dark energy component. We observe the smallest uncertainty of $\sum m_\nu$ in the bin between $10\leq z\leq 100$. The sensitivity of the neutrino mass bound starts to weaken at a similar redshift as the onset of dark energy domination (see below).  The degeneracy between the sum of neutrino masses and the dark energy density might play an important role here. \\

\begin{table*}[t]
\begin{tabular}{l|c|c|c|c}
Dataset & CMB & CMB + CMBL & CMB + CMBL + BAO & CMB + CMBL + BAO + SN\\
\hline
\hline 
$\sum m_\nu(z\leq 0.5)$ (eV) & 4.86 & 3.07 & 1.73 & 1.13 \\
$\sum m_\nu(0.5 \leq z\leq 3)$ (eV) & 3.19 & 2.85 & 0.51 & 0.42 \\
$\sum m_\nu(3 \leq z\leq 10)$ (eV) & 1.64 & 1.63 & 0.40 & 0.37 \\
$\sum m_\nu(10 \leq z\leq 100)$ (eV) & 0.47 & 0.50 & 0.20 & 0.19 \\
$\sum m_\nu(100 \leq z\leq 1100)$ (eV) & 0.39 & 0.37 & 0.33 & 0.32 \\
$\sum m_\nu(z\geq 1100)$ (eV) & 0.44 & 0.42 & 0.41 & 0.40 \\
\hline
\hline 
\end{tabular} 
\caption{\label{tab:results} 95\% upper limits on the amplitudes of the neutrino mass at different redshifts reconstructed with a binned parametrization and using different data combinations.}
\end{table*}

We note in Fig.~\ref{fig:mnu_lambda} that in all cases there is little correlation between the individual neutrino mass amplitudes. We also studied how the inclusion of growth rate measurements obtained with redshift-space distortions (RSD)~\cite{Alam:2016hwk} would affect our results presented here. RSD can be used to probe the growth rate $f(k,z)$ at different scales and redshifts~\cite{Percival:2008sh}. The scale-dependent feature in this quantity is a source of information for neutrino masses~\cite{Hernandez:2016xci,Villaescusa-Navarro:2017mfx,2011MNRAS.418..346M,Boyle:2017lzt}. Including the growth rate measurements, we found a decrease of the bound of $\sum m_\nu$ of 38\% in the lowest redshift bin and only a marginal improvement in other bins. The exact modeling of RSD is under active development and therefore we present our final results without including RSD.

\subsection{Reconstruction with splines}
Next we reconstruct the neutrino mass sum with linear splines and variable knots (see Sec.~\ref{sec:methods}).
We investigate at which redshift the sensitivity on $\sum m_\nu$ starts to weaken by allowing the knots $z_1$ and $z_2$ to vary as well. We show our results for this reconstruction on the right-hand side of Fig.~\ref{fig:mnuz_data}. We also present our results in Table~\ref{tab:results_splines}. For the final data combination we find 
\begin{align*}
    \sum m_\nu(z=0)&<1.46 \text{ eV}\\
    \sum m_\nu(z=1100)&<0.53 \text{ eV}
    \hspace{3em}\text{(95\% CL)} .
\end{align*}

\begin{table*}[t]
\begin{tabular}{l|c|c|c|c}
Dataset & CMB & CMB + CMBL & CMB + CMBL + BAO & CMB + CMBL + BAO + SN \\
\hline
\hline 
$\sum m_\nu(z=0)$ (eV) & 4.75 & 2.91 & 1.43 & 1.46 \\
$\sum m_\nu(z=0.5)$ (eV) & 3.50 & 2.74 & 2.31 & 0.76 \\
$\sum m_\nu(z=3)$ (eV) & 2.11 & 1.84 & 0.18 & 0.18\\
$\sum m_\nu(z=10)$ (eV) & 1.07 & 1.11 & 0.16 & 0.15 \\
$\sum m_\nu(z=100)$ (eV) & 0.69 & 0.47 & 0.17 & 0.18 \\
$\sum m_\nu(z=1100)$ (eV) & 0.52 & 0.50 & 0.53 & 0.53 \\
\hline
\hline 
\end{tabular} 
\caption{\label{tab:results_splines}95\% upper limits on the amplitudes of the neutrino mass at different redshifts reconstructed with linear splines and variable knots and using different data combinations.}
\end{table*}

\begin{figure}[t!]
    \centering
    \includegraphics[width=0.9\columnwidth]{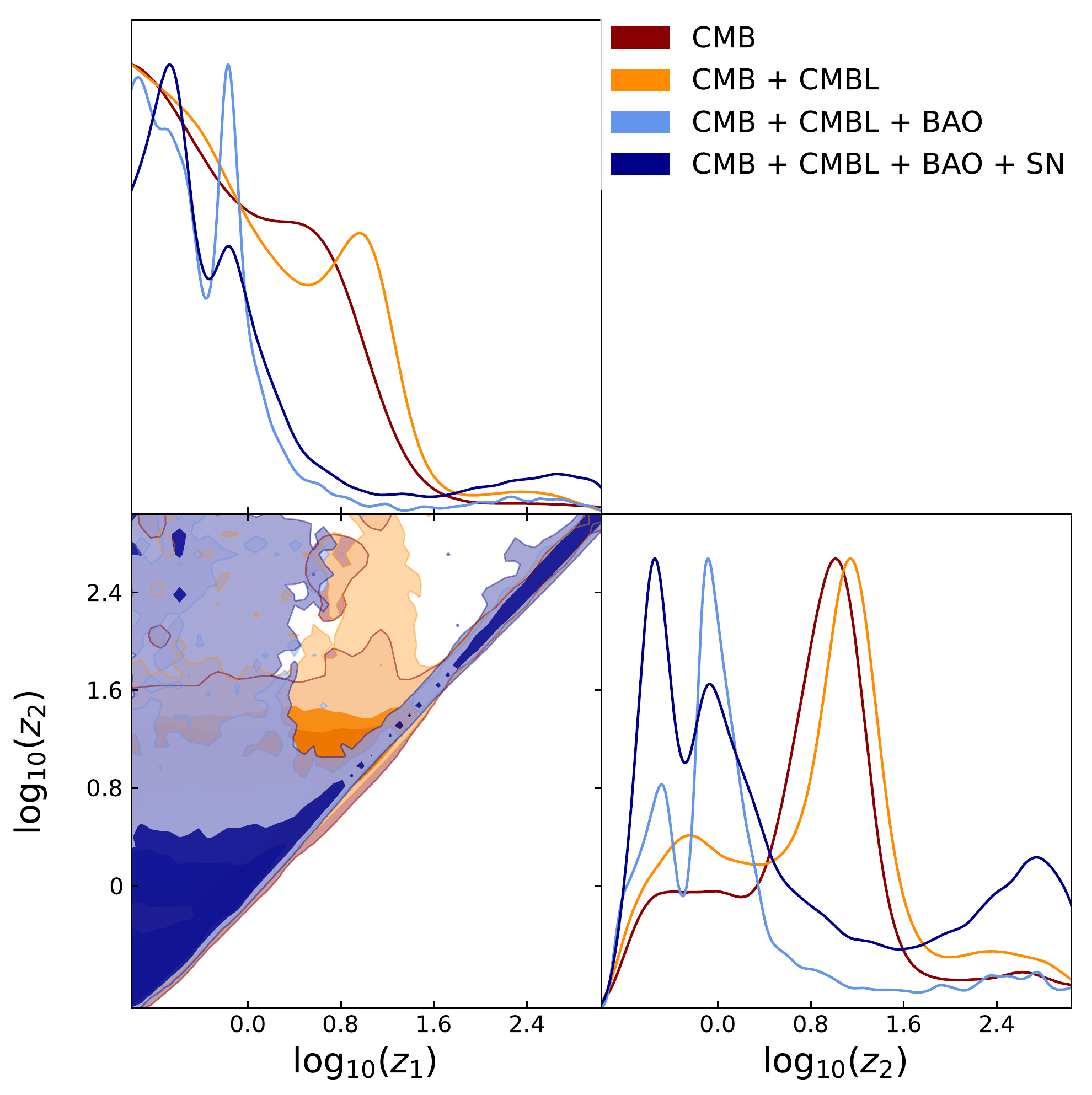}
    \caption{The results for the change points $z_1$ and $z_2$ from the reconstruction with linear splines and variable knots.}
    \label{fig:change}
    \end{figure} 
    
\begin{figure*}[ht!]
    \centering
    \includegraphics[width=1.3\columnwidth]{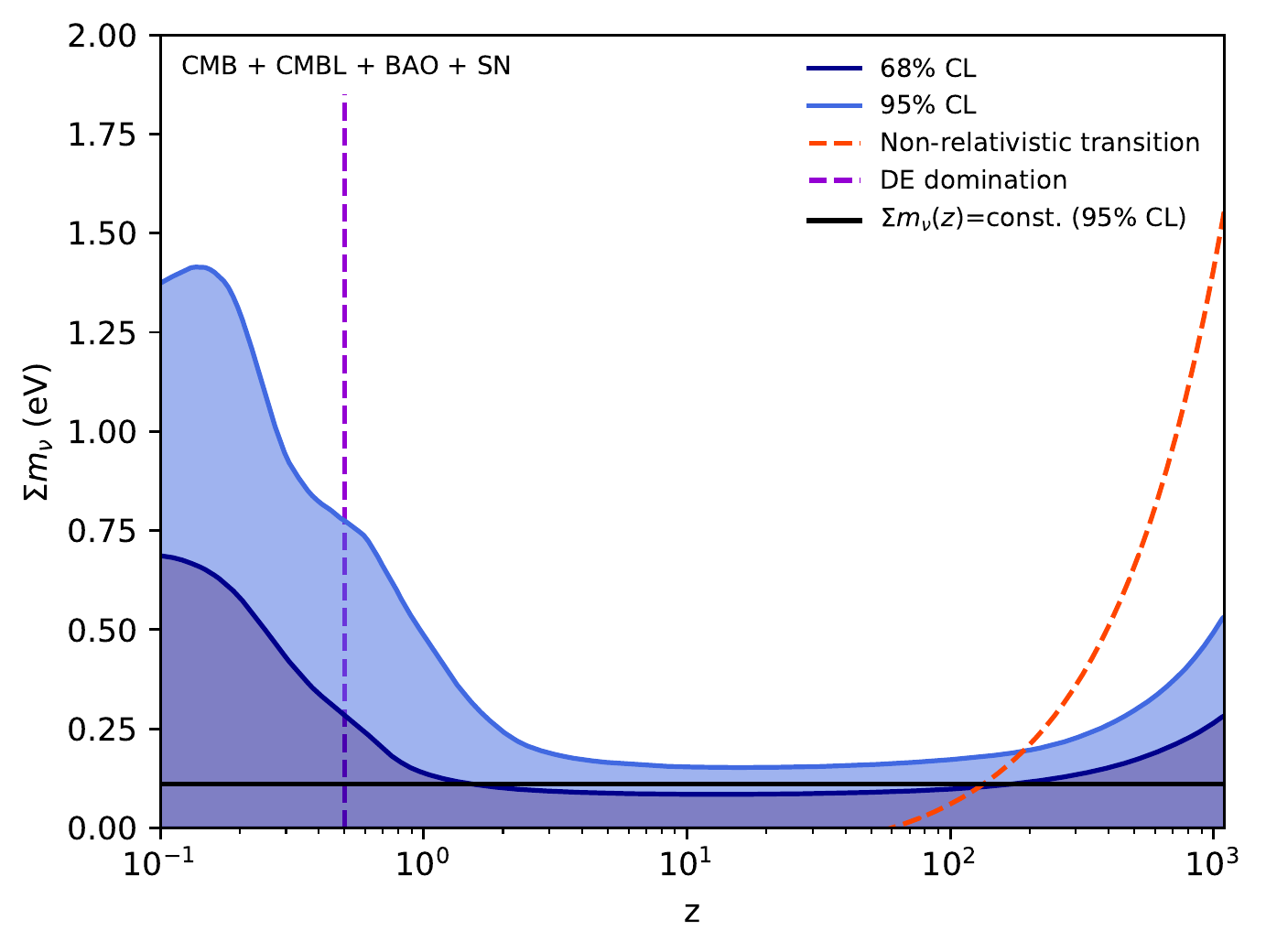}
    \caption{Final result: Neutrino mass limits as function of redshift obtained from \textit{Planck} 2018 CMB temperature, polarization, and lensing data, BAO data from BOSS DR12, 6dF, and MGS, eBOSS DR14 quasars, and Lyman-alpha, and SN data from Pantheon 18 (zoom of the lower right-hand panel in Fig.~\ref{fig:mnuz_data}). The onset of dark energy domination is denoted with a dashed violet line. The dashed red line shows $\sum m_\nu$ in the case when the neutrino with the largest neutrino mass becomes nonrelativistic at redshift $z$, assuming a normal neutrino mass hierarchy.}
    \label{fig:obs}
\end{figure*} 

In general, when comparing the results from the two different reconstruction methods, we note that the constraints for $\sum m_\nu$ at different redshifts are  similar. There are small differences, in particular for $\sum m_\nu(z=0)$, because in the binned reconstruction $\sum m_\nu$ is fixed between the bin margins.
We also note that, as expected, we obtain smooth curves for the reconstruction with linear splines. This is due to the fact that the positions of the knots $z_1$ and $z_2$ are free parameters in the analysis, smoothing the posterior means and credible bands. Compared to the binned reconstruction, we observe that the credible bands of the reconstructed curve via splines with variable knots exhibit more features despite the same number of free parameters. 

For constant neutrino masses in time, we would, in principle, expect flat posteriors for the two knots $z_1$ and $z_2$, as the position of the knots does not matter when fitting a constant function. However, even in that case, the posterior for $z_1$ and $z_2$ can still vary because of possible degeneracies between the neutrino mass sum and other cosmological parameters and because of the different amount of data points at different redshifts, resulting in local changes of the uncertainty of the neutrino mass sum. 

In Fig.~\ref{fig:change}, we see that there is no clear preference for the position of the knots $z_1$ and $z_2$. For the position of the first knot (change point) we find $z_1<0.89$ (68\% CL) for the full data combination. This limit is close to the onset of dark energy domination at $z\sim 0.5$~\cite{Velten:2014nra}. The preference for a small value for the position of $z_1$ is the most pronounced for the final data combination. 

The position of the second knot, $z_2$, is less constrained, but also moves toward lower redshifts when including more data. Whereas the first knot, $z_1$, mostly is needed to model the increase of the uncertainty of $\sum m_\nu(z)$ at low redshifts, the second knot, $z_2$, determines the width of the regime where the uncertainty of $\sum m_\nu(z)$ is the smallest. 

\subsection{Implications for neutrino mass decay models}
\label{sec:decay}

In this section, we note that our \textit{a priori} model-independent reconstruction can constrain specific models that predict new physics beyond the Standard Model of Particle Physics. In particular, as discussed in Sec.~\ref{sec:models}, it has been proposed that neutrinos can decay into dark radiation in the late Universe. To apply our constraints to these models, we need to carefully distinguish two cases. First, in the scenario in which neutrinos decay while they are still relativistic, the anisotropic stress changes before the decay~\cite{Barenboim:2020vrr}. This makes it hard to directly transfer our bounds to such models, because these models change the neutrino properties both for $z<z_{\rm nr}$ and for $z>z_{\rm nr}$. If instead neutrinos decay after they become nonrelativistic, all cosmological neutrino properties for $z>z_{\rm nr}$ are the same as in the standard case. Thus, any change of neutrino properties for $z<z_{\rm nr}$ due to the decay will not affect the properties for $z>z_{\rm nr}$. In particular, any constraints on cosmological parameters before the decay, such as $\sum m_\nu$ or $\omega_\nu h^2$, will be unaltered in the model compared to the standard $\Lambda$CDM model with constant-mass, nondecaying neutrinos. This implies that our constraints for $z>z_{\rm nr}$ can be directly transferred to nonrelativistic neutrino decay models.

Our constraint is shown in our final result plot, Fig.~\ref{fig:obs}, which will be discussed in more detail in the next section.  In this plot, the red dashed line is the neutrino mass sum as a function of the redshift $z_{\rm nr}$ at which the largest neutrino mass eigenstate $m_{\nu_3}$ becomes nonrelativistic~\cite{Ichikawa:2004zi},
\begin{equation}
    1+z_\mathrm{nr}\sim \frac{2\,m_{\nu_3}}{\text{meV}},
\end{equation}
assuming a normal neutrino mass hierarchy. The smaller neutrino mass eigenstates will become nonrelativistic later and thus can be neglected in this curve. Taking the intersection of the red dashed line and our result curve, we find a limit of
\begin{equation}
    \sum m_\nu<0.21 \text{ eV}
    \hspace{3em}\text{(95\% CL)} 
    \label{eq:decay}
\end{equation}
for $z>z_{\rm nr}$. For the nonrelativistic decay scenario, we bound $\sum m_\nu$ with our result for $\sum m_\nu(z=203)$. Note that $103<z_\mathrm{nr}<203$ are the redshifts at which the individual neutrinos would become nonrelativistic if they had a combined mass of $\sum m_\nu=0.21$~eV (95\% CL) (see Fig.~\ref{fig:obs}). 

Using \textit{Planck} 2015 data, it has been previously shown that neutrino decay models can relax the cosmological neutrino mass bound to $\sum m_\nu \lesssim 0.9$~eV for the non-relativistic decay scenario~\cite{Escudero2019,Chacko:2019nej,Chacko:2020hmh,Escudero:2020ped}. Note that the cosmological impact of the resulting dark radiation is negligible in these models, as the neutrinos  decay during the matter domination era.
Compared to the earlier results in~\cite{Escudero2019,Chacko:2019nej,Chacko:2020hmh,Escudero:2020ped}, our bound in Eq.~\eqref{eq:decay} might be tightened due to two different effects. First, we use more recent CMB data from the latest 2018 \textit{Planck} release~\cite{Aghanim:2018eyx}. Second, the redshift of the nonrelativistic neutrino transition decreases with decreasing neutrino mass, such that the bound gets tighter for later neutrino decays (see Fig.~4). We also note that the neutrino mass bound of $\sum m_\nu \lesssim 0.9$~eV~\cite{Escudero2019,Chacko:2019nej,Chacko:2020hmh,Escudero:2020ped} relied on simplified assumptions for the equations of motions, as noted in Ref.~\cite{Barenboim:2020vrr}. This might have an impact on this model-specific mass bound but does not alter our results due to the reasons discussed above.



The limits extracted above do not include all fundamental assumptions of the nonrelativistic decay models, e.g., a decrease in the total neutrino mass. In principle, it is possible that the wider limits that we find at low redshifts could influence the limits at high redshifts. To check the robustness of our estimates, we extended the extraction to include a theoretical prior
on the neutrino mass sum with $\sum m_{\nu,z_i}\leq\sum m_{\nu,z_j}$ for $z_i\leq z_j$.Using this prior, we found an upper limit of $\sum m_\nu<0.26$ eV (95\% CL) for the nonrelativistic decay scenario. 
This shows that the neutrino mass bound for the decay models is stable with respect to a potential late-time change of the neutrino mass sum.

\section{Discussion and Conclusion}\label{sec:conclusion}

In this paper, we reconstructed the cosmological neutrino mass sum as a function of redshift. This reconstruction was model independent, such that no specific mass model or interaction with the dark energy or dark radiation sector was assumed.
Our final result is shown in Fig.~\ref{fig:obs}. The figure shows the reconstructed neutrino mass sum for the full data combination of 
CMB temperature, polarization and lensing, combined with BAO and SN data. We highlight the redshift when dark energy starts to dominate. We also highlight the redshift $z_{\rm nr}$ at which the largest neutrino mass eigenstate becomes nonrelativistic, in order to constrain models~\cite{Escudero2019,Chacko:2019nej,Chacko:2020hmh,Escudero:2020ped} that predict neutrino decay after this nonrelativistic transition (see Sec.~\ref{sec:results} and below). \\

Our result is consistent with neutrino masses that are constant in time, as predicted by the Standard Model of Cosmology. We observe a small increase of the bound on $\sum m_\nu$ at high redshifts ($z\gtrsim 500$), as well as a large increase of this bound at low redshifts coinciding with the onset of dark energy domination. The large mass bound at low redshifts could be explained in two different ways.

On the one hand, it could simply arise due to the strong degeneracy between $\Omega_\Lambda$ and $\sum m_\nu$ at low redshifts, which is shown in Fig.~\ref{fig:mnu_lambda}. We have explicitly tested this scenario by imposing a strong prior on $\Omega_\Lambda=0.69\pm 0.01$. In that case, the constraint on $\Sigma m_{\nu}(z=0)$ decreased by 42\% to $\sum m_{\nu}(z=0)=0.84$ eV, which shows an impact from correlations with dark energy but is still larger than standard neutrino mass bounds. This latter effect might be due to the large number of additional free parameters needed for the reconstruction. 

On the other hand, the large mass bound at low redshifts could be explained by new physics. It has been a longstanding puzzle why the energy scales of neutrino masses and dark energy are close ($m_\nu\sim\sqrt[4]{\rho_\Lambda}\sim{\rm meV}$) but far away from all other known fundamental energy scales, such as the Higgs or Planck scales. If cosmological data allow for larger neutrino masses in the late Universe, $0\leq z\leq 1$ (such as suggested in Refs.~\cite{Brookfield_2006,Brookfield2006,Koksbang:2017rux,Lorenz:2018fzb,Muir:2020puy,Battye:2013xqa,Wyman:2013lza,Beutler:2014yhv,Poulin:2018zxs}), an intriguing theoretical connection to the dark energy sector could appear~\cite{Fardon:2003eh,Bjaelde:2007ki,Ayaita:2014una,Mandal:2019kkv,DAmico:2018hgc,Dvali:2016uhn}. In particular, our results agree with the insights from previous work presented in Ref.~\cite{Lorenz:2018fzb}. In that study, a specific model was assumed that predicted the late neutrino mass generation in the Universe arising from
a supercooled phase transition.
We note that our results also show the same trend recently obtained in Ref.~\cite{Muir:2020puy} by the Dark Energy Survey collaboration, where a higher neutrino mass and a corresponding low $\Omega_\Lambda$ was found. \\

The analysis presented in this paper could be extended in different ways. On the theory side, it would be interesting to reconstruct $\sum m_\nu(z)$ fully implementing potential interactions with the dark energy or the dark radiation sector (see below). In particular, a coupling between the dark energy and the neutrino sectors would affect the dark energy perturbations, leading to more features in the model and additional ways to decrease the degeneracy between dark energy and massive neutrinos. 

Our approach is conservative in the modeling of the neutrino mass but not necessarily in the constraints that we obtain: we choose the simplest parametrization of the mass, without any model-specific assumptions, but the constraints could be weakened or strengthened when model-specific physics is introduced. For example, the neutrino mass bounds could become much weaker for modified gravity models~\cite{Bellomo:2016xhl,Hagstotz:2019gsv,Hagstotz2019} or when introducing a nontrivial coupling between the neutrino and dark energy sectors, for example in the model considered in Ref.~\cite{Lorenz:2018fzb}. On the other hand, the bounds are expected to become stronger in neutrino decay models when taking into account neutrino perturbations~\cite{Escudero2019,Chacko:2019nej,Chacko:2020hmh,Escudero:2020ped}. Such features could be captured by a more general reconstruction of the neutrino mass including nontrivial couplings to different sectors.

On the methodology side, one could model the number of knots or bins with a hyperparameter \cite{DimatteoGenoveseKass01} in order to allow the data to decide on the number of knots. To speed up computations, this can be fitted with empirical Bayes, where the hyperparameters are chosen as the maximizers of the marginalized likelihood~\cite{2012PhRvD..86h3001S,Kern_2017}. The exact number of knots or bins can impact the results by under- or overfitting the reconstructed function.

On the data side, it would be interesting to reconstruct $\sum m_\nu$ including more large scale structure data, in particular galaxy clustering, cosmic shear, and cross-correlations of either the $\ell$ISW effect and galaxy clustering or of CMB lensing and tracers of the large scale structure. These additional measurements of the suppression of the matter power spectrum would also be an additional way to break the degeneracy with dark energy~(see, e.g.~\cite{2010MNRAS.405..168L,Mishra-Sharma:2018ykh,Brinckmann:2018owf,Boyle:2018rva,1836511,Pearson:2013iha,Giusarma:2018jei,Yu:2018tem,Lesgourgues:2007ix,Xu:2016jns,Bayer:2021iyb}). To fully exploit these data will however require careful modeling of the nonlinear scales (see, e.g. ~\cite{PhysRevLett.100.191301,Brandbyge_2010,Lesgourgues:2009am,Wong:2011ip,2012MNRAS.420.2551B,Adamek:2017uiq,Villaescusa-Navarro:2019bje,Mead:2020vgs,Nascimento:2021wwz,Bose:2021mkz}), and of potential systematics, such as the scale-dependent galaxy bias in the presence of massive neutrinos~\cite{Castorina:2015bma,Raccanelli:2017kht,Vagnozzi:2018pwo,Chiang:2018laa,Xu:2020fyg}. 

Finally, we comment on the implications of our results for particle physics models and experiments. In particular, our study strengthens the previously reported cosmological mass bound $\sum m_\nu <0.9$~eV (95\% CL) of nonrelativistic neutrino mass decay scenarios~\cite{Escudero2019,Chacko:2019nej,Chacko:2020hmh,Escudero:2020ped} to $\sum m_\nu<0.21$ eV (95\% CL). We note that the constraints are still weaker than for constant neutrino masses in time, because decay scenarios are insensitive to late-time cosmological data from BAO and SN. However, our study pushes the neutrino mass bounds of such models below the sensitivity of the KATRIN experiment. This implies that a neutrino mass discovery at KATRIN would hint toward other models, such as models predicting post-recombination neutrino mass generation and subsequent relic neutrino annihilation, such as proposed in Ref.~\cite{Dvali:2016uhn}. The cosmological disappearance of the neutrino mass parameter might also be explained in the context of modified gravity~\cite{Bellomo:2016xhl,Hagstotz:2019gsv,Hagstotz2019}, but this degeneracy between massive neutrino and modify gravity effects will be broken by future surveys, such as Euclid~\cite{Hagstotz2019}. 

\section*{Acknowledgments} 
We thank Alexandre R\'{e}fr\'{e}gier, Thejs Brinckmann, Sunny Vagnozzi, Pedro Machado, Roni Harnik, Rapha\"{e}l Sgier, Jessie Muir, Yvonne Wong and Neal Dalal for helpful discussions and/or comments on the draft. We also acknowledge helpful discussions at the Cosmology from Home Conference in 2020. CSL acknowledges financial support by Alexandre R\'{e}fr\'{e}gier's Cosmology research group at ETH Zürich. Research at Perimeter Institute is supported in part by the Government of Canada through the Department of Innovation, Science and Industry Canada and by the Province of Ontario through the Ministry of Colleges and Universities. EC acknowledges support from the STFC Ernest Rutherford Fellowship ST/M004856/2 and STFC Consolidated Grant ST/S00033X/1, and from the European Research Council (ERC) under the European Union’s Horizon 2020 research and innovation programme (Grant agreement No. 849169). ML has been funded in part by ETH Foundations of Data Science (ETH-FDS). Parts of this research was conducted using the computer cluster Euler at ETH Zürich. 

\bibliography{ref}

\appendix
\onecolumngrid

\end{document}